# Development and Understanding of Nb$_3$Sn films for radiofrequency applications through a sample-host 9-cell cavity

T. Spina[1], B. M. Tennis[1], J. Lee[1,2], D. N. Seidman[2,3] S. Posen[1]

[1]*Fermi National Accelerator Laboratory, Batavia, IL, 60510, USA*
[2]*Department of Materials Science and Engineering, Northwestern University, Evanston, IL, 60208, USA*
[3]*Northwestern University Center for Atom-Probe Tomography (NUCAPT), Evanston, IL, 60208, USA*

ABSTRACT

Nb$_3$Sn is a promising advanced material under development for superconducting radiofrequency (SRF) cavities. Past efforts have been focused primarily on small development-scale cavities, but large, often multi-celled cavities, are needed for particle accelerator applications. In this work, we report on successful Nb$_3$Sn coatings on Nb in a 1 m-long 9-cell Nb sample-host cavity at Fermilab. The geometry of the first coating with only one Sn source made it possible to study the influence of Sn flux on the microstructure. Based on these results, we postulate a connection between recently observed anomalously large thin grains and uncovered niobium spots observed in the past by other authors [Trenikhina 2018]. A phenomenological model to explain how these anomalously large grains could form is proposed. This model is invoked to provide possible explanations for literature results from several groups and to guide key process parameters to achieve uniform vapor-diffusion coatings, when applied to complex structures as the multi-cell cavity under study.

## 1. INTRODUCTION

There are ongoing efforts at institutions worldwide towards developing Nb$_3$Sn coatings for superconducting radiofrequency (SRF) cavities [Stimmel 1978; Allen 1983; Mayer 1986]. Due to its high critical temperature, T$_c$ (18 K), Nb$_3$Sn cavities can have smaller surface resistances at a given temperature than standard Nb cavities. It follows that high quality Nb$_3$Sn coatings would make it possible to decrease significantly cryogenic costs for SRF-based accelerators. Its high superheating field may also lead to higher ultimate accelerating fields [Posen 2015].

Vapor-diffusion [Saur 1972; Hillenbrand 1977; Peiniger 1988] is the Nb$_3$Sn coating technique that has so far produced the highest performing films, yielding very uniform, homogeneous and stoichiometric thin layers $\sim 2 \mu m$. It involves the reaction at high temperatures of a niobium substrate, such as an SRF cavity, with Sn vapor, typically in a vacuum furnace.

Because Nb$_3$Sn has still been in fairly early stages of development compared to niobium, most efforts have been focused on small cavity geometries that are not practical for particle accelerators. This has been useful for research purposes, but now that performances have been demonstrated that would be useful for applications if they could be scaled up [Posen 2017; Pudasaini 2019c], there is increasing motivation to explore coatings on larger practical cavities. A typical geometry is a multicell elliptical cavity, used for accelerating particles with a velocity close to the speed of light. An early effort to coat a 5-cell elliptical cavity was performed at Wuppertal University in the 1980s, which involved a two-step coating cycle with a total time of 60 hours. The performance of the cavity was limited to only low accelerating fields <5 MV/m [Peiniger 1989].

Using the current methods that achieve strong performance on single-cell cavities and scaling them up to multicell cavities is not necessarily straightforward due to the nature of the process, which involves thermal evaporation of Sn vapor from a heated source. Due to the complex geometry of SRF cavities, substrate surfaces occupy a variety of angles and distances from the source, as well as areas outside of the line-of-sight, issues which have complicated other coating processes as well (e.g. Nb/Cu).

Furthermore, RF performance degradation is difficult to diagnose on coated cavities. Even if problems can be localized to a region of the cavity, microscopy typically cannot be performed except by destructive techniques, i.e., cutting coupons from the cavity.

These factors motivate the use, in this study, of a sample-host cavity; a cavity that matches the desired geometry but has holes in its walls to hold samples. The cavity would never undergo RF testing but would rather be used to understand how different process parameters alter the final surface in different regions of the cavity. The cavity could be used for parameter optimization in conjunction with RF tests of actual cavities with the corresponding geometry, to help provide an understanding of how the parameters are affecting the local microstructure and connect that to the overall RF performance. Sample-host cavities have been used in the past to study other SRF processing techniques [Cooper 2013; Chouhan 2019; Lee 2017]

From a materials science perspective, the Nb$_3$Sn vapor diffusion process is not well understood. There are a number of empirical observations that have been made in the past that have led to the coating procedures currently used. An important example is the observation by Hillenbrand *et al.* [1977] of the inclination for apparent *uncovered niobium spots* to occur in the Nb$_3$Sn layer (also described in the literature as problems with nucleation). The structure and composition of these spots was not uncovered at the time (though some investigations suggested they may include a thin Sn-containing layer rather than only uncovered niobium), but Hillenbrand *et al.* [1980a] did develop and publish methods to prevent these spots from forming. These methods include pre-anodizing the niobium substrate, providing a high flux of Sn vapor early in the coating process, and the use of high vapor pressure Sn halides. Some or all of these methods are typically used in modern vapor diffusion coatings, but their mechanism for helping improve layer quality has not been clear.

More recent studies have been focused on the investigation of anomalously large thin grains (so-called *patchy regions*). These were first studied after being directly correlated with RF degradation in a cavity [Posen 2015b] and were analyzed in studies at Fermilab, Cornell, and JLab [Trenikhina 2018; Posen 2017; Pudasaini 2017]. Compared to typical grains size (diameter ~1µm), these grains are very large (diameter as large as ~100µm) [Trenikhina 2018]. Their thinness is consistent with their large size—bulk diffusion of Sn through grains is expected to be substantially slower than grain boundary diffusion according to theoretical calculations [Besson 2007; Fedorov 1979; Landolt 1990]. Far from the grain boundaries towards the center of the grain, there would be no fast way for Sn to diffuse to the Nb$_3$Sn/Nb interface to grow additional Nb$_3$Sn. The thickness of these grains (observed down to ~100 nm) is small enough that RF fields are expected to penetrate to intermediate Nb/Sn phases below the Nb$_3$Sn layer, which are expected to have poor superconducting properties (lower $T_c$ and superheating field) [Godeke 2006].

Without samples from Hillenbrand et al. [Hillenbrand 1977] it hasn't been possible to directly compare the apparent uncovered Nb regions they had observed in the past to the recently studied anomalously large thin grain regions. However, there are reasons to suspect that they are the same features. Both have a similar visual appearance by eye, as shiny spots on the cavity surface. Just as Hillenbrand et al. observed that pre-anodization helps to reduce the appearance of apparent uncovered regions, Hall *et al.* [2017] observed similar effects on anomalously thin grains. Similarly, just as observed by Hillenbrand, the use of Sn halides and a high flux from the Sn source help to reduce the appearance of apparent uncovered regions. Recently, Pudasaini *et al.* [2019b] observed similar effects on anomalously thin grains. Developing a model to help understand how these features evolved may be helpful for understanding past results, and it may also help to guide successful coatings of complex structures.

In this work, the Nb$_3$Sn vapor diffusion coating treatment has been applied to a 9-cell sample-host cavity to optimize the coating process and to study the effect of Sn flux on microstructure. Two coating treatments have been performed, with one and two Sn sources. Because of this multi-cell geometry, the coating treatment with just one Sn source represents a useful experimental set-up to study the evolution of microstructural feature formation with a varying amount of Sn flux delivered in each cell.

2. EXPERIMENTAL SET-UP

*2.1 Samples and assembly*

The sample-host cavity started as a Nb 9-cell cavity with a significant manufacturing defect. The cavity had never been treated or tested but was sent directly to the Fermilab machine shop, where holes were added in the outer region of each cell (the equators) and in the regions between consecutive cells (the irises), Figure 1. The cavity houses 17 samples, and they are numbered by the cell they are in (i.e., #1.0 is the sample located on the equator of cell 1, closest to the input coupler port of the cavity), or else the cells that they are in between (i.e., #1.5 is the sample located on the iris between cells 1 and 2), Figure 2.

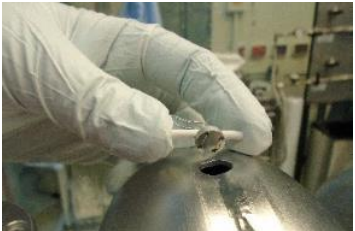

**Figure 1:** Host samples under study. Note the hole in the cavity.

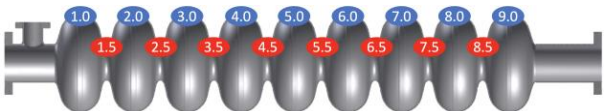

**Figure 2:** Two types of sample locations on the 1.3GHz 9-cell cavity: samples on the irises (in red) and on the equators (in blue).

All NbTi parts were removed from the cavity for ease of coating, and simple Nb flanges were welded on to its ends. The cavity was given a buffered chemical polish (BCP) after machining. Before each coating, the cavity was BCP'd, degreased, ultrasonically cleaned, and high-pressure water rinsed. Additionally, each of the machined holes was filled with a 1-cm diam. 3 mm thick bulk Nb sample, Figures 1 and 2. These samples were machined from corners of RRR=300 niobium sheets used to fabricate cavity cells. They were electropolished (EP'd) prior to coating.

For coating, the cavity is placed in Fermilab's $Nb_3Sn$ coating system, consisting of a horizontal furnace that has been modified to support a coating chamber, Figure 3. Heaters are located on one or both ends of the cavity (two heaters are shown in Figure 3). The heaters contain crucibles with Sn and $SnCl_2$ which are evaporated during the coating procedure. R-type thermocouples are placed in each heater to control the temperature (two thermocouples in each for redundancy). Note that because the thermocouples are not located directly where the Sn is, there is some offset between the reading and the Sn temperature.

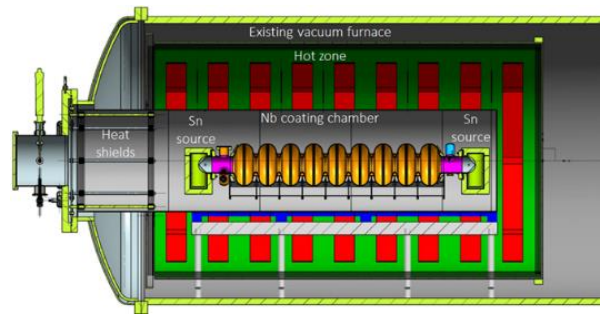

**Figure 3:** Horizontal furnace for $Nb_3Sn$ multicell cavity coating at Fermilab.

Two coating treatments have been performed and compared: 1-Sn source and 2-Sn sources, Figure 4. For the coating with only 1-Sn source, the other end of the cavity was open to the coating chamber. Neither the cavity nor the samples were pre-anodized for the coating with one Sn source, but they were pre-anodized for the coating with two Sn sources.

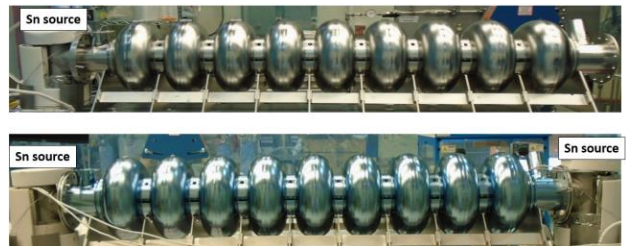

**Figure 4:** Set-up for Nb3Sn 9-cell cavity: 1-Sn source (top); 2-Sn sources (bottom) – SHC9

Surface pre-anodization treatment is used to grow the oxide layer of the niobium substrate before coating, utilizing electrolytic methods. The use of pre-anodization and its effects was examined by Hillenbrand et al. [Hillenbrand 1980a], finding that it helped to prevent the growth of "patchy [...] optically blank areas." Hall *et al.* [2017], compared the growth of Nb$_3$Sn layers of pre-anodized samples and non-pre-anodized samples, demonstrating that surface pre-anodization can suppress the formation of anomalous large thin film regions. It was found, however, that high quality Nb$_3$Sn films could be formed without pre-anodization [Posen 2015]. The mechanism by which pre-anodization prevents the growth of optically blank/anomalously large thin grains has been unclear.

### 2.2 *A15 phase diagram: region of interest for Nb$_3$Sn SRF cavity*

Nb$_3$Sn is an intermetallic compound belonging to a class of materials known as A15s, most of them stable at the stoichiometric composition, A$_3$B. As reported by Flükiger [1982], the type of formation of the A15 phase in a system A$_{1-\beta}$B$_\beta$ (where β is the B composition) is essentially governed by the B element; at thermodynamic equilibrium, it can form the compounds Nb$_{1-\beta}$Sn$_\beta$ (0.18<β≤0.25), Nb$_6$Sn$_5$ and NbSn$_2$, according to the generally accepted binary phase diagram by Charlesworth et al. [Matthias 1954]. In particular, the region that corresponds to 0.18<β≤0.25 is the portion of the phase diagram that is of greatest interest for fabrication of an SRF surface. Devantay *et al.* [1981] found that the superconducting and normal state properties depend critically on the composition β [Muller 1980] and, for example, T$_c$ decreases significantly below 24 at.% Sn. It follows that to obtain high performance Nb$_3$Sn cavities, it is important that the material produced not only sit in this range but also has a composition very close to stoichiometry (25at.%Sn). Thus, the A15 phase diagram region of interest for SRF applications has to be restricted further to the smaller range 24 ÷ 25 at.% Sn.

### 2.3 *Multicell vapor-diffusion coating procedure*

The Nb$_3$Sn phase can be formed either above 930°C in the presence of a Sn – Nb melt, as during the vapor-diffusion coating process adopted in this work or below this temperature by solid-state reactions between Nb and Nb$_6$Sn$_5$ or NbSn$_2$. As displayed in Figures 5 and 6, both coatings performed in this study have similar features in their temperature profiles. Five stages can be identified: (i) degassing; (ii) nucleation; (iii) ramp-up; (iv) coating; (v) annealing; and (vi) cooldown. An extensive description of these stages can be found in [Posen 2017], while herein a comparative description of the two-coating treatments is discussed.

For both coatings under study, during the degassing stage the temperature in the furnace, T$_{Furnace}$, is held between 100°C and 200°C under vacuum, to remove residual moisture, which may have been introduced while the furnace was open. This stage lasts ~4 hours.

Then, T$_{Furnace}$ is raised to 500°C for 5 h and at this temperature, the SnCl$_2$ is expected to evaporate fully, creating Sn sites on the surface of the substrate.

Following this, there is a ramp-up from 500°C to the coating temperature, during which the Sn source heater(s) are activated, raising the temperature of the Sn source, T$_{Sn}$, to ~100 ÷ 300°C higher than T$_{Furnace}$, which the niobium substrate temperature is expected to follow fairly closely.

During the coating stage that lasts around 2 h, the cavity is maintained at a constant temperature above 950°C, which corresponds to the temperature at which low-T$_c$ phases (Nb$_6$Sn$_5$ and NbSn$_2$) are unfavorable thermodynamically. Typically, T$_{Furnace}$ is chosen to be ~1100°C and T$_{Sn}$ ~100 to 200°C higher. The Sn heater permits the flux to be controlled independently of the Sn diffusion rate in the Nb$_3$Sn layer, which is strongly influenced by the substrate temperature.

After coating, to reduce the rate of Sn arriving at the surface, the secondary heater is turned off. The furnace is maintained at coating temperature for ~2 additional hours to allow any excess of pure Sn at the surface of the cavity to diffuse into the layer or evaporate.

Several of the coating methods discussed above (pre-anodization, use of Sn halides, higher temperature Sn source than cavity) were recommended by Hillenbrand *et al.* [1980a]. They reported that the use of these methods helps to achieve reproducible results free of apparent uncovered niobium spots.

They report that they cannot explain what happens in this procedure, but they suggest that "nucleation may become more homogeneous with a high Sn supply at the niobium surface." They note that having the Sn source hotter than the substrate provides a high Sn flux and the thicker oxide of the anodized surface prevents contact of Sn with metallic niobium until it dissolves—thereby giving more time for additional Sn to arrive before the reaction starts in earnest. They recommended the use of Sn halides ($SnF_2$ or $SnCl_2$), which have a high vapor pressure at a relatively low temperature, to avoid having to generate a thermal gradient between the Sn source and the substrate [Hillenbrand 1977].

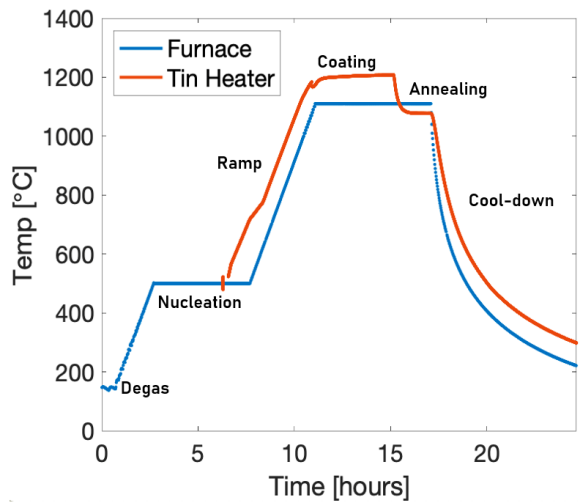

**Figure 5:** Temperature profile during coating: 1-Sn source

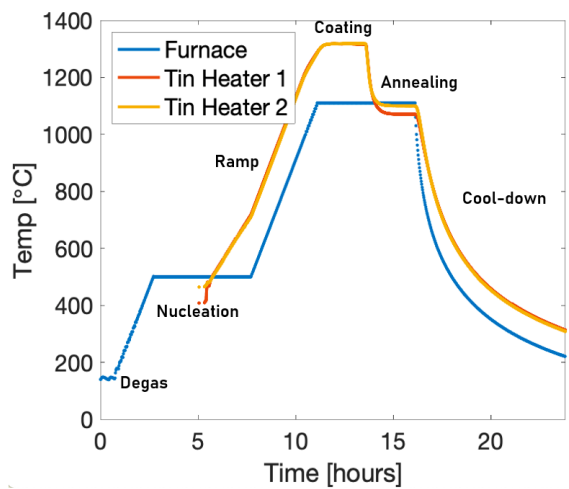

**Figure 6:** Temperature profile during coating: 2-Sn sources

As is observed from the temperature profiles in Figures 5 and 6, the maximum Sn heater temperature, $T_{Sn}$, was significantly higher in the 2-Sn sources coating (~1300°C) than in 1-Sn source (~1200°C). The cause for the low temperature in the first coating was later discovered to be that the heater had fallen off of the can containing the Sn crucible, apparently due to a failed support wire. Additional supports were welded to the ends of the cavity frame to hold the weight of the heaters in future coatings, which appeared to have been sufficient to prevent the same issue from occurring in the 2-Sn sources coating.

For 1-Sn source coating, the Sn evaporation rate is expected to be significantly smaller than usual due to the low $T_{Sn}$. An attempt was made to offset this by increasing the duration of the coating by one hour, but this was not very successful as can be seen in Table 1.

| Coating/Source | Sn Initial | Sn Evaporated |
|---|---|---|
| 1-Sn source coating, source A | 5.0 g | 1.6 g |
| 2-Sn sources coating, source A | 5.0 g | 5.0 g |
| 2-Sn sources coating, source B | 5.0 g | 2.5 g |

**Table 1**: Initial Sn mass and Sn mass evaporated during the two coatings from each Sn source. Sources A and B are located close to cells 1 and 9, respectively. Values were obtained by weighting the crucibles before and after coating.

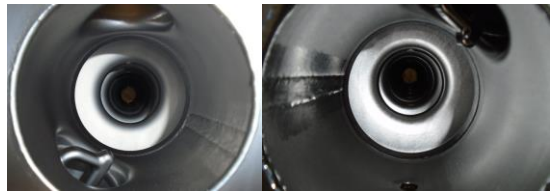

**Figure 7** 1-Sn source inside views from both sides. The left image was closest to the Sn source and shows a matte $Nb_3Sn$ appearance. By eye, the other side looks like uncoated Nb.

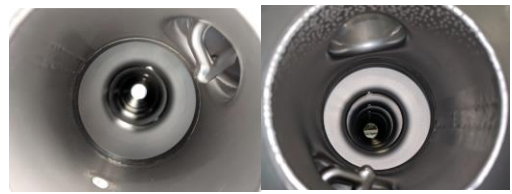

**Figure 8:** 2-Sn sources, inside views from both sides.

Figures 7 and 8 display images looking into the sample host cavity after each coating, and some initial observations can be made prior to discussing the microscopic measurements on samples. In the 1-Sn source coating, on the coupler side where the Sn source was, the cavity looks to be coated with Nb$_3$Sn. On the other side, furthest from the Sn source, the cavity resembles bare Nb. In the 2-Sn sources coating, both sides look like matte Nb$_3$Sn, with the exception of some Sn spots very close to the flange, probably due to an excess of Sn flux from the source.

3. MEASUREMENTS AND RESULTS

Samples extracted from each cell were characterized systematically using microscopy measurements. In particular, SEM/EDS measurements were performed at Fermilab and FIB/HAADF-TEM/EBSD at Northwestern University.

*3.1 Electron Microscopy:*

Scanning electron microscopy (SEM), energy dispersive x-ray spectroscopy (EDS), transmission electron microscopy (TEM) and electron backscatter diffraction (EBSD) measurements were performed on each sample, Figure 1, to study the microstructural evolution and composition as a function of Sn flux.

SEM images from the samples located at different positions after the 1-Sn source coating are displayed in Figure 9. Going from cell #1.0 (closest to the Sn source) to cell #9.0 (farthest from Sn source) it is possible to distinguish three different regions.

In particular, the four samples located in the first two cells (#1.0; #1.5; #2.0; #2.5), closest to the Sn source, present a fairly homogeneous coating (same coating thickness on the whole sample surface) with the composition being very close to stoichiometry (24.6at.%Sn). Samples from cells three and four display an intermediate behavior: only some regions display a homogeneous coating with a composition close to stoichiometry, while other regions show patchy, anomalously thin grains. Beginning from cell five until the end (far from Sn-source), nearly all areas were covered with patchy anomalously thin grains.

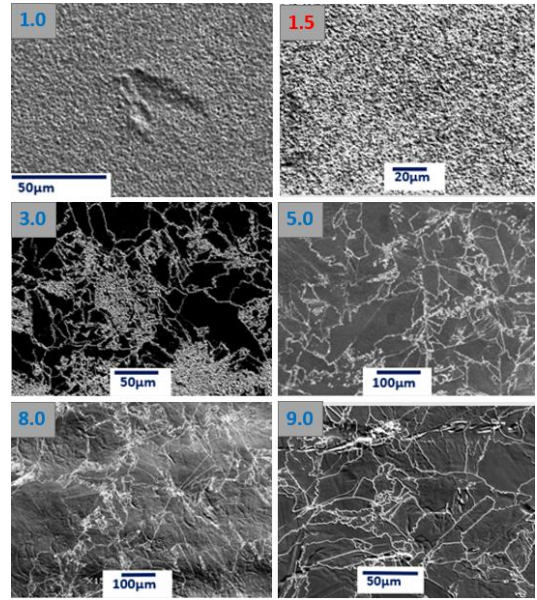

**Figure 9:** SEM: 1-Sn source, 25 kV

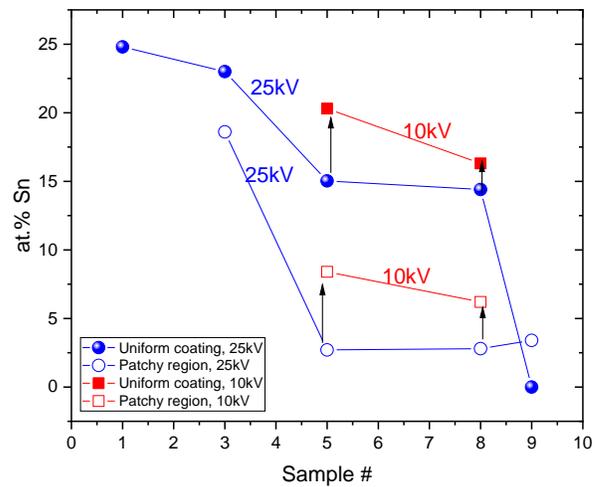

**Figure 10**: EDS results of samples from 1- Sn source coating. The Sn source is closest to sample #1.0 in cell 1.

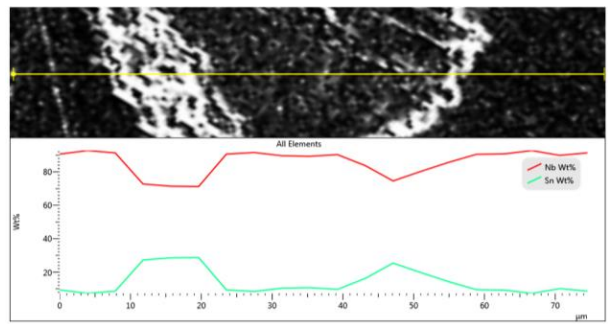

**Figure 11**: EDS line scan at 25 kV: patchy anomalously thin grain in sample #5.0.

A summary of the EDS composition values as a function of Sn-distance after the 1-Sn source treatment is shown in Figure 10. To study possible changes in composition as a function of coating thickness, the EDS measurements were performed at two different electron-beam energies, hence exciting small (10 kV, less than 1 µm$^3$) and large (25 kV, more than 1 µm$^3$) volumes and comparing the effects at different depths. The deeper penetrating 25 kV beam measures a higher Nb content, suggesting that it is passing through a relatively thin Nb$_3$Sn layer into the Nb bulk. More evidence is found from a 25 kV EDS line scan across *patchy regions*, Figure 11: Nb$_3$Sn thinner layer at the center with a thicker Nb$_3$Sn layer displayed around it.

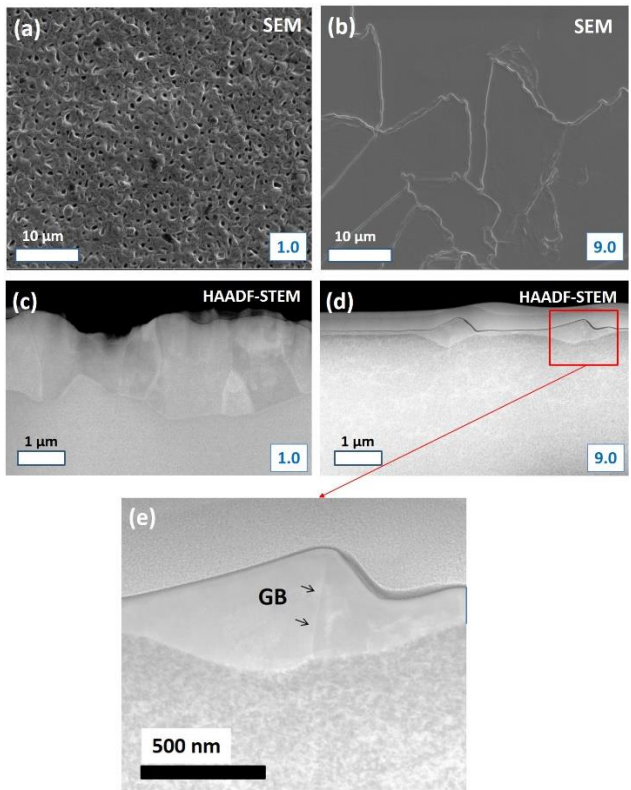

**Figure 12:** High resolution SEM and cross-sectional HAADF-STEM images of 1-Sn source coating samples #1.0 (a,c) and #9.0 (b,d). Extremely thin grains are displayed in the sample #9, and (e) the film thickness is ~5 times thicker near GBs than the grain interior indicating that GBs are the main path for Sn-diffusion and supply for the growth of Nb$_3$Sn grains.

Additional SEM measurements with higher resolution were performed using a FEI Quanta FEG 650 at Northwestern University on samples #1.0 (close to Sn source) and #9.0 (far from Sn source) and are displayed in Fig. 12(a) and 12(b), respectively. The higher resolution reveals a porous structure, similar to that reported in [Pudasaini 2016].

Cross-sectional images by high-angle annular dark-field (HAADF) imaging STEM measurements were performed for studying the effect of high- and low-Sn content on the Nb$_3$Sn layer thickness. As shown in Figure 12(c), sample #1.0 shows a very homogeneous ~2 µm thick Nb$_3$Sn layer, while sample #9.0, Figure 12(d), is characterized by large thin Nb$_3$Sn grains resembling *patchy regions*. Regions near the grain boundaries are thicker by a factor 5 than those inside grains, Fig. 12(e). This observation is a direct demonstration of the different mechanisms of Sn diffusion in Nb$_3$Sn and supports strongly the idea that Sn diffusion through grain boundaries is much faster than bulk diffusion. An EBSD map performed on sample #9.0, Figure 13, reveals that these *patchy regions* have large lateral grain diameters, with sides in many cases larger than 50 µm.

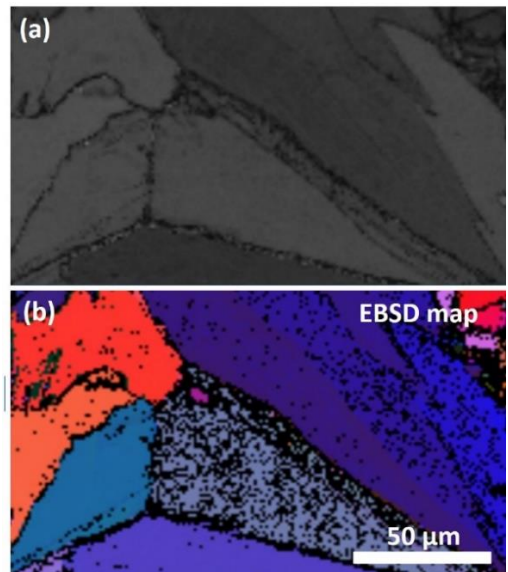

**Figure 13:** (a) Band contrast and (b) EBSD map on the patchy regions indicates that the grain diameter is more than ~50 µm.

Very different results were obtained on the samples from the 2-Sn sources coating. For this treatment, the samples and cavity were pre-anodized, the Sn flux per source was substantially higher (Table 1), and the two Sn sources were directed at different ends of the cavity for maximum coverage. As shown from SEM observations, Figure 14, all samples from this treatment had a more homogeneous coating along the length and no patchy anomalously thin-grain regions were observed.

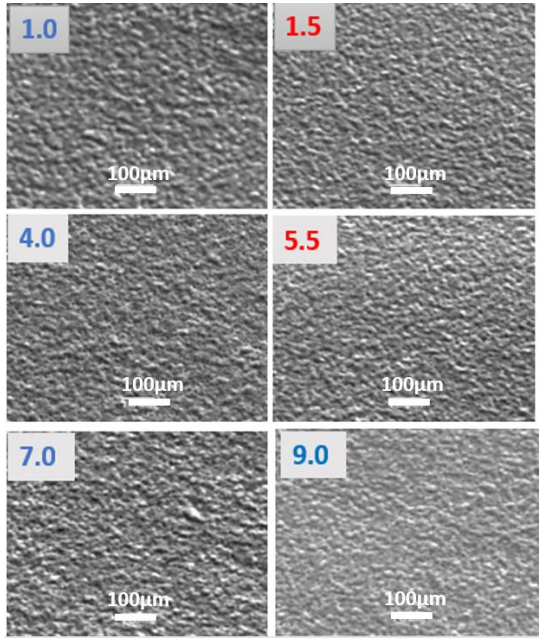

**Figure 14:** SEM: 2-Sn sources: very homogeneous coating along the 9-cell cavity length.

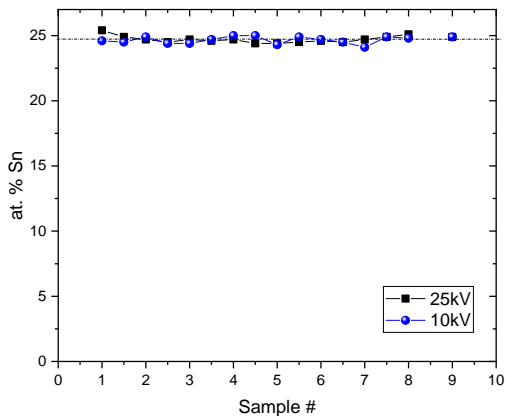

**Figure 15:** EDS: 2-Sn sources: stoichiometric coating along the 9-cell cavity length.

In Figure 15, composition values from EDS analyses as a function of the electron beam energy (10 kV and 25 kV) reveals almost no variation in the microstructures proving that the 2-Sn sources treatment can produce great uniformity of film thickness along the length of multi-cell structures.

This finding is also supported by TEM observations, Figure 16, showing a constant ~2 µm thick $Nb_3Sn$ layer despite the different locations of the samples cut out. Thickness changes along the 1 m length 9-cell cavity was estimated to be about ~46%, Figure 16(a).

In Figure 17, samples from the two coatings are compared using high-resolution SEM images. The porous appearance that affected the 1-Sn source coating is not displayed in the 2-Sn sources coating. The cause of the porous microstructure is unclear. One possible speculation may be due to the Sn flux being insufficient, but not low enough to cause large thin grain patchy regions.

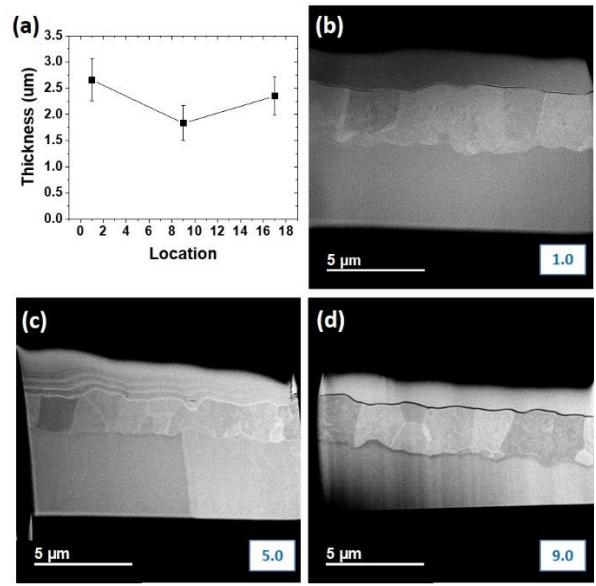

**Figure 16** (a) Film thicknesses (more than 2 µm) of $Nb_3Sn$ coatings at various locations (1, 9, 17) in a host cavity demonstrate that the thickness is reasonably uniform along the 9-call cavity. There is an ~46% variation in the thickness at different locations, 1 to 9, in the host cavity.

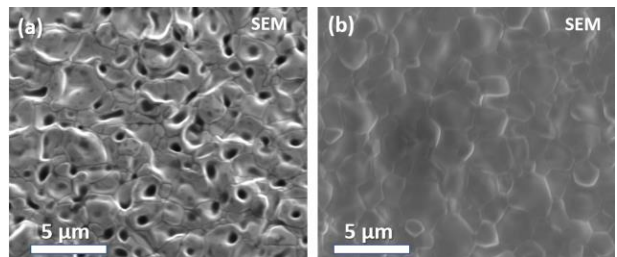

**Figure 17** Higher resolution SEM comparison: (a) Coating with one Sn source (sample #1.0); (b) two Sn sources (sample #9.0)

Deeper analyses of the $Nb_3Sn$ grain boundaries (GBs) in the 1-Sn and 2-Sn sources coating were performed and are displayed in Figure 18 (see details of procedure in [Lee 2020]). For the 2-Sn sources coating, Sn segregation was not observed in sample #1.0, but it was observed in sample #9.0 from the 2-Sn sources coating. This may be related to the fact that Sn source A, closest to sample #1.0, had evaporated all of its Sn, while Sn source B, closest to sample #9.0, had Sn remaining, meaning that some Sn would have continued to evaporate during the annealing step, possibly leading to additional Sn accumulation in the GBs. In future coatings, less Sn could be used in source B.

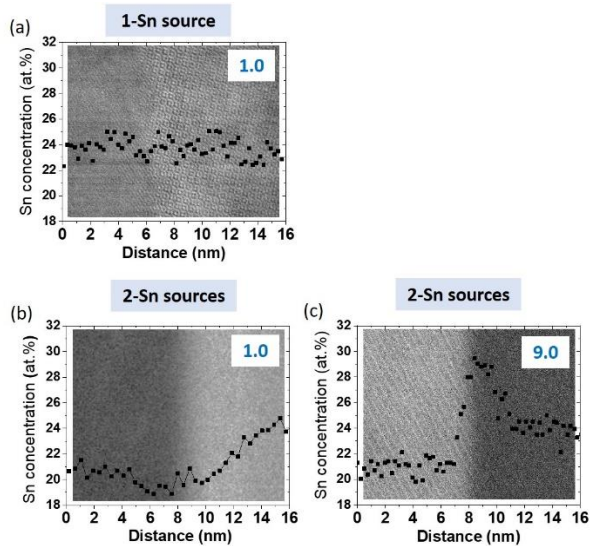

**Figure 18**: STEM-EDS measurements of concentration profiles across a grain boundary in $Nb_3Sn$ coatings prepared by: (a) 1-Sn source; and (b,c) 2-Sn sources. Sample #1 is selected for 1-Sn source and samples #1 and #9 are selected for 2-Sn source.

4. DISCUSSION

While there were already strong indications from previous studies that anomalously large thin grains can form as a result of a low Sn flux [Pudasaini 2019b], the experimental findings on the samples from the 1-Sn source coating act as further evidence, and they provide an indication of how the morphology changes as the Sn flux changes. There is a clear trend that further away from the Sn source, where the Sn flux is expected to be smaller, more and more of the area is covered with *patchy regions* (see images in Figure 9 and EDS measurements in Figure 10). With these measurements contributing to a growing body of evidence, it seems very likely that the apparent uncovered Nb spots observed by Hillenbrand [1980] were indeed anomalously large thin grains.

Generally, the presence extremely thin $Nb_3Sn$ regions are anticipated to be problematic for SRF performance. But the fact that these appear specifically as *large* thin grains may pose a worse problem: once large thin grains have formed, due to the reduced areal fraction of GBs in $Nb_3Sn$ coatings, Sn diffusion is limited, preventing layers from thickening significantly. Indeed, GBs in $Nb_3Sn$ coatings play pivotal roles in the growth of $Nb_3Sn$ layers and the formation of large thin grains at an early stage of the coating process may act as a persistent "poison" to the $Nb_3Sn$ layer during the remainder of the coating process. The growth process is irreversible and the microstructure in an early stage affects the final stage of the coating.

This has important implications for the $Nb_3Sn$ coating process, especially for large structures, such as multi-cell cavities. It means that a long coating process with a relatively low Sn flux (e.g. this may apply to the 60-hour-long coating of the Wuppertal 5-cell cavity) is likely to result in patchy regions. A short process with a high Sn vapor pressure to all regions of the cavity is more likely to produce a successful film. In other words, the final coating is affected not just by the integrated dose of Sn to the surface, but also by the local dose rate during the coating process.

This is somewhat counterintuitive. Because of the natural feedback provided by the phase diagram of $Nb_3Sn$, it seems advantageous to perform a slow, low Sn flux procedure. If one provides a relatively small Sn flux from the source, the Sn vapor may be concentrated at first on the first cell of the cavity. But with time, the first cell should eventually become saturated (as the $Nb_3Sn$ layer becomes thicker, the overall rate of Sn diffusion to the surface decreases, and the rate of Sn arrival outpaces it). Any extra Sn sticking to the surface may become liquid (but not form Sn-rich poor-$T_c$ phases at this temperature) and reevaporate. Then the second cell should have more Sn vapor available, until it to becomes saturated, and so on, until the entire cavity is covered with a $Nb_3Sn$ layer. Also, this Sn-vapor diffusion process seems like it could coat an arbitrary cavity geometry, but the evidence suggests it would be prone to the formation of patchy regions. To prevent this, it appears that a high Sn flux needs to be provided to all surfaces of the cavity early in the process.

Next, why do patchy regions form easily in case of small-Sn flux and why does a high-Sn flux help to prevent them? It could be related to the formation of texturing in case of small-Sn-flux during the coating process. A relevant observation was reported by Lee *et al.* [2019] that in areas where *patchy regions* were found, there is evidence of certain orientation relationships between the Nb$_3$Sn layer and the Nb substrate. The images in Figure A1 in the Appendix support this idea. The existence of texture in large-thin grains on a single-grain Nb (110) substrate suggests that there are favorable interfacial free energies between Nb$_3$Sn and Nb for the local orientation relationship in the anomalously large thin grains. There could be some additional effect of surface free energy on the Nb$_3$Sn coating but the correlation between the inverse pole figures of Nb$_3$Sn and Nb indicate that the interfacial free energy may be the primary cause of the texture in the Nb$_3$Sn coating in Figure A1 [Thompson 2000].

Based on the observations of texturing, Sn flux dependence, and other factors (including dependence of having a thick oxide layer, which may act as a barrier between the Nb and the Sn to permit for a significant amount of Sn to build up on the surface before meeting the Nb layer), two possible models are proposed below for how *patchy regions* form:

1) **Nucleation dominated process**: early in the growth process, Nb$_3$Sn nuclei form on the surface.
    a. If the *Sn flux is low*, a sparse network of nuclei form. These nuclei coarsen as the Sn vapor arrives. As the nuclei meet one another, nuclei that have an energetically favorable orientation (either relative to the substrate or the surface) may influence nearby nuclei to rotate into the same orientation.
    b. If the *Sn flux is high*, a dense network of nuclei form. Even though some nuclei may randomly have an energetically favorable orientation. Once the Nb$_3$Sn nuclei impinge forming a dense network of GBs, the Nb$_3$Sn nuclei (grains), have a difficult time rotating to form stable Nb$_3$Sn/Nb interfaces.
2) **Time dominated process**: the Sn flux determines the duration for which the Nb$_3$Sn layer exists, but it is very thin.
    a. If the *Sn flux is low*, there will be a long duration in which the thickness of the Nb$_3$Sn layer is very small, meaning that the energy associated with the Nb/Nb$_3$Sn interfaces will contribute with a large fraction to the overall energy of a grain [Thompson 2000]. Over many minutes or hours with a low Sn flux, grains with a favorable interfacial energy may have a chance to grow laterally, forming high aspect ratio thin grains.
    b. If the *Sn flux is high*, the film may thicken quickly enough to make the interfacial free energy less of a dominant factor in the overall free energy.

Other mechanisms are, of course, possible as well. Mechanisms like those above would explain why the empirically developed methods (pre-anodization, use of Sn halides, higher temperature Sn source than the cavity) help to prevent the formation of uncovered niobium areas. As is described in detail in the following paragraphs, this connection can be used to offer potential explanations of some of the results reported in the literature and can also guide efforts to maintain Nb$_3$Sn layer quality when coating large structures.

The models above could explain why *patchy regions* were observed in the ERL1-5 cavity at Cornell, which is the first report of anomalously large thin grains [Trenikhina 2018]. That cavity, ERL1-5, was the first cavity coated at Cornell, and it had poor RF performance, which was correlated to the patchy regions by means of a temperature mapping system. For its coating procedure, during the ramp up to the coating temperature, the Sn heater had only minimal power applied, so that the Sn source and cavity temperature were very close during the ramp to coating temperature. For later, better performing cavity coatings, the Sn source was ramped with a temperature 100 to 150°C higher than that of the cavity, based on the recommendation of Hillenbrand et al. [Posen 2015a]. This would help to explain why substantial *patchy regions* were found in ERL1-5 but typically not in later Cornell cavities. It is also noted that at an attempt was made to coat ERL1-5 a second time, without removing the first coating; similar to the procedure used on the Wuppertal 5-cell cavity. Based on the above discussion, if anomalously large thin grains had already formed, it seems unlikely that an additional coating step would improve the coating quality and uniformity.

The connection between Sn flux and *patchy regions* may explain another significant feature of coatings in the literature—why Wuppertal cavities experienced a strong Q-slope degradation starting at 5 MV/m, and Cornell's did not [Peiniger 1988; Muller G 1996; Posen 2014]. Proposed reasons in the past include the possibility that an extra-long annealing step may have played a key role [Posen 2014]. But another explanation should be considered: it is possible that the Wuppertal coating procedure resulted in anomalously large thin grains, and the Cornell procedure was able to avoid this because of a higher Sn flux.

In the Cornell coating setup, the cavity was left open to the coating chamber [Posen 2015a] which may have allowed for a higher Sn flux to the surface without adverse effects. The Wuppertal setup on the other hand has a lid that covers the top of the cavity, closing the volume [Peiniger 1988]. Early coatings at Fermilab tried to adopt a similar procedure to that used at Wuppertal, but it led to an apparent condensation of Sn droplets on the cavity surface during coating, when a similar evaporation rate, to that used at Cornell, was employed, ~2 g over 3 hours [Hall 2017]. This is displayed in Figure A2 in the Appendix. It seems likely that Cornell's open cavity geometry permitted a large Sn flux to be used without causing condensation on the cavity surface—once the cavity surface became saturated, excess vapor would instead be transferred to the coating chamber.

Similarly, Fermilab now employs an open cavity geometry, whenever possible, to permit a high Sn-flux without condensation by utilizing the coating chamber and cold region of the furnace as a Sn "sink." JLab has also used a procedure that permits a high Sn flux; in their case not by opening the cavity to the coating chamber, but by connecting the main cavity to be tested to a secondary "sacrificial" cavity to absorb the excess Sn flux [Pudasaini 2019c].

The understanding, based on the discussion herein, was applied when developing the coating parameters for the 2-Sn source coating procedure. For that coating procedure, both Sn sources were heated as high as possible to have as high a Sn flux as possible; the cavity and samples were pre-anodized; $SnCl_2$ was used to create a high vapor pressure early in the coating procedure; during ramp-up, the Sn source temperature was maintained higher than that of the cavity temperature; and the coupler port, transmitted power port, and higher order mode ports of the cavity were open to try to prevent condensation. Some condensation still appeared to have occurred near 1-Sn source, Figure 7, but this is in the region, where the RF fields are relatively small. Overall, this approach seems to have been positive. The coating with two Sn sources successfully produced samples with near-ideal composition over the entire length of the 9-cell cavity, with good thickness, and good stoichiometry, as shown in Figure 15.

Future work will include simulations of vapor flow in the 9-cell cavity to understand how the vapor pressure affects the coating. This is a complex situation involving changes in diffusion rate, sticking coefficients, and re-evaporation rate over the surface as a function of time. Additionally, the flow is expected to be in the transition region of the Knudsen number between molecular flow and continuum flow; Sn vapor equilibrium mean free path at 1200°C ~5 cm [Peiniger 1989] and cavity dimension ~10 cm, further complicating flow simulations.

Soon after these experiments were performed on the 9-cell sample host cavity, a real 9-cell cavity was coated successfully and achieved significantly improved performance compared to the Wuppertal 5-cell cavity, see [Posen 2019] for details.

Contemporaneously JLab's efforts to coat 5-cell cavities resulted in similarly successful coatings [Eremeev 2015].

5. CONCLUSIONS

Two $Nb_3Sn$ vapor diffusion processes on 9-cell cavities have been compared and discussed. The 1-Sn source coating provided a unique opportunity to study the microstructure as a function of Sn flux due to the geometry of the cavity, showing clearly that the formation of anomalously large thin grains is more likely when the Sn flux is small. Higher vapor Sn flux supplied during the 2-Sn source coating led to a uniform ~2 µm thick $Nb_3Sn$ coating layer without *patchy regions*. A model was presented on how these regions form, which was used to provide possible explanations for previous results of $Nb_3Sn$ vapor diffusion coatings. The model was used to guide coatings of large structures, e.g., 9-cell cavities, including recommendations for utilizing a Sn "sink" to permit a high Sn flux without condensation. Using this information, the coating with two Sn-sources led to a highly uniform $Nb_3Sn$ multi-cell coating with only normal regions being observed in all the samples investigated.

ACKNOWLEDGEMENTS

This manuscript has been authored by Fermi Research Alliance, LLC under Contract No. DE-AC02-07CH11359 with the U.S. Department of Energy, Office of Science, Office of High Energy Physics and supported by a DOE Early Career Award. This work made use of the EPIC, Keck-II, and/or SPID facilities of Northwestern University's NU*ANCE* Center, which received support from the Soft and Hybrid Nanotechnology Experimental (SHyNE) Resource (NSF ECCS-1542205); the MRSEC program (NSF DMR-1121262) at the Materials Research Center; the International Institute for Nanotechnology (IIN); the Keck Foundation; and the State of Illinois, through the IIN.

tin on niobium for vapor diffusion coating of $Nb_3Sn$," Supercond. Sci. Technol. 32 045008.

**[Pudasaini 2019c]** Pudasaini U. et al., 2019, "Recent Results From $Nb_3Sn$ Single Cell Cavities Coated at Jefferson Lab," Proceedings of The Nineteenth International Conference on RF Superconductivity, Dresden, Germany, MOP0183, doi:10.18429/JACoW-SRF2019-MOP018.

**[Saur 1962]** Saur E. and Wurm J., 1962, "Preparation und Supraleitungseigenschaften von Niobdrahtprobe n mit $Nb_3Sn$-Uberzug" Die Naturwissenschaften 49, Issue 6, pp. 127-128.

**[Stimmell 1978]** Stimmel, 1978, "Microwave superconductivity of $Nb_3Sn$", PhD Thesis Cornell University

**[Thompsom 2000]** Thompsom C. V., 2000, "Structure evolution during processing of polycrystalline films" Ann. Rev. Mater. Sci.

**[Trenikhina 2018]** Trenikhina Y., Posen S., Romanenko A., Sardela M., Zuo J-M, Hall D. L., Liepe M., 2018, "Performance defining properties of $Nb_3Sn$ coating in SRF cavities", Supercond. Sci. Technol. 32 015004 (13pp)

APPENDIX

Figure A1 shows a $Nb_3Sn$ coating on a *single-grain* Nb substrate with surface orientation close to (110). It was coated with extremely low Sn-flux in an effort to intentionally induce patchy regions and study their texture. EBSD analysis of this patchy region illustrates that for the large (~50 μm), thin $Nb_3Sn$ grains, several specific orientations appear frequently. This texture is clearly visible in the inverse pole figures (IPF). Because the coating was performed on a single-grain Nb substrate, if it consistent with having a correlation between the formation of large thin grains and them having specific, energetically favorable orientation relationships.

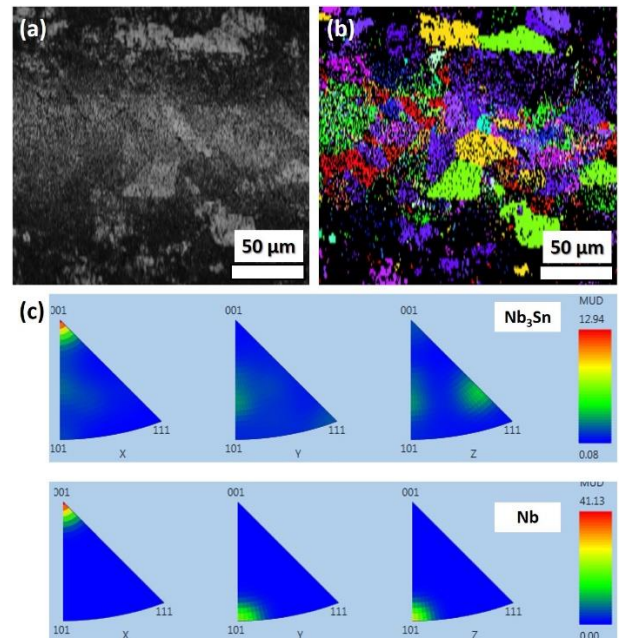

**Figure A1:** $Nb_3Sn$ coating on a single-grain Nb sample. Patchy regions with large thin grains formed. (a) band contrast image; (b) EBSD orientation map on z-direction. (c) Inverse Pole Figure (IPF) for the EBSD maps of the $Nb_3Sn$ coating and Nb substrate shows that the $Nb_3Sn$ grains are textured on the Nb substrate.

Figure A2 shows images from an early coating at Fermilab, in which the cavity volume was mostly closed, rather than being open to the coating chamber. In this coating, spots appeared on areas that had the most direct line of sight to the Sn source. The interpretation is that these spots were condensed Sn droplets that later converted to $Nb_3Sn$ (based on EDS measurements of their composition).

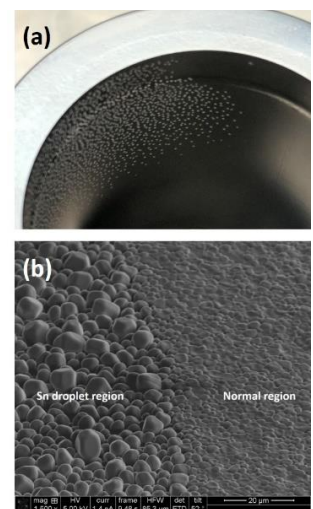

**Figure A2**: (a) Visible spots in line of sight to the tin source that were potentially caused by Sn droplets during the coating process; and (b) SEM image showing atypical $Nb_3Sn$ grain growth in the droplet area.

14